\documentclass[12pt]{article}
\pdfoutput=1

\usepackage{amsmath}
\usepackage{amsfonts}
\usepackage{amscd}
\usepackage{amsthm}
\usepackage{setspace}

\usepackage{graphicx}
\usepackage{authblk}
\usepackage{caption}
\usepackage{ytableau}
\usepackage{mathtools}

\def\Z{\mathbb{Z}}
\def\Q{\mathbb{Q}}

\def\C{\mathbb{C}}
\def\P{\mathbb{P}}

\def\til{\tilde}

\begin{document}

\begin{titlepage}

\begin{flushright}
KEK-TH 2007
\end{flushright}

\vskip 1cm

\begin{center}

{\large Structure of stable degeneration of K3 surfaces \\into pairs of rational elliptic surfaces}

\vskip 1.2cm

Yusuke Kimura$^1$
\vskip 0.4cm
{\it $^1$KEK Theory Center, Institute of Particle and Nuclear Studies, KEK, \\ 1-1 Oho, Tsukuba, Ibaraki 305-0801, Japan}
\vskip 0.4cm
E-mail: kimurayu@post.kek.jp

\vskip 1.5cm
\abstract{F-theory/heterotic duality is formulated in the stable degeneration limit of a K3 fibration on the F-theory side. In this note, we analyze the structure of the stable degeneration limit. We discuss whether stable degeneration exists for pairs of rational elliptic surfaces. We demonstrate that, when two rational elliptic surfaces have an identical complex structure, stable degeneration always exists. We provide an equation that systematically describes the stable degeneration of a K3 surface into a pair of isomorphic rational elliptic surfaces. When two rational elliptic surfaces have different complex structures, whether their sum glued along a smooth fiber admits deformation to a K3 surface can be determined by studying the structure of the K3 lattice. We investigate the lattice theoretic condition to determine whether a deformation to a K3 surface exists for pairs of extremal rational elliptic surfaces. In addition, we discuss the configurations of singular fibers under stable degeneration. 
\par The sum of two isomorphic rational elliptic surfaces glued together admits a deformation to a K3 surface, the singular fibers of which are twice that of the rational elliptic surface. For special situations, singular fibers of the resulting K3 surface collide and they are enhanced to a fiber of another type. Some K3 surfaces become attractive in these situations. We determine the complex structures and the Weierstrass forms of these attractive K3 surfaces. We also deduce the gauge groups in F-theory compactifications on these attractive K3 surfaces times a K3. $E_6$, $E_7$, $E_8$, $SU(5)$, and $SO(10)$ gauge groups arise in these compactifications.}  

\end{center}
\end{titlepage}

\tableofcontents
\section{Introduction}
\label{sec1}
\par The F-theory approach to particle physics model building has several advantages. It naturally realizes SU(5) grand unified theories with matter in spinor representations of SO(10). In contrast to D-brane models, there is no difficulty in generating up-type Yukawa couplings. Furthermore, it can evade the problem of weakly coupled heterotic string theory addressed in \cite{Witten}. Recent studies on F-theory model building \cite{DW1, BHV1, BHV2, DW2} have emphasized the use of local models. However, in order to address the issue of gravity such as inflation, a global model of compactification need to be considered eventually. In particular, many insights can be gained by the duality between heterotic string and F-theory \cite{Vaf, MV1, MV2, Sen, FMW}, which states the equivalence between the former compactified on an elliptically fibered CY $n$-fold and the latter on a K3 fibered CY $(n+1)$-fold in the stable degeneration limit \cite{FMW, AM} \footnote{For recent discussion of the stable degeneration limit of F-theory and F-theory/heterotic duality, see, for example, \cite{AHK, BKW, BKL, CGKPS, MizTan}.}. The aim of the present paper is to develop a systematic study of the process of stable degeneration of a K3 surface for a pair of rational elliptic surfaces.
\par K3 surface stably degenerates into two rational elliptic surfaces in two distinct ways: \\
i)K3 surface splits into two rational elliptic surfaces with an identical complex structure \\
ii)K3 surface splits into two rational elliptic surfaces with different complex structures. \\
We discuss these cases separately, in section \ref{sec2} and section \ref{sec3}, respectively. 
\par We demonstrate in section \ref{subsec2.1} that, when a pair of rational elliptic surfaces are isomorphic, stable degeneration can be described by a systematic equation. We analyze the geometry of stable degeneration of the first kind i) using this equation. We determine that given any pair of isomorphic rational elliptic surfaces, there is some K3 surface that stably degenerates into the pair.  
\par However, it is considerably difficult to describe stable degeneration by an equation when two rational elliptic surfaces have different complex structures. Moreover, a pair of non-isomorphic rational elliptic surfaces glued together along a smooth fiber does not necessarily admit a deformation to a K3 surface. Complex structures and configurations of singular fibers are classified for a specific class of rational elliptic surfaces, called extremal rational elliptic surfaces. We focus on the extremal rational elliptic surfaces to analyze the stable degeneration of the second kind ii). We determine whether stable degeneration exists for pairs of these rational elliptic surfaces. We use the lattice theoretic approach to analyze this process.
\par  We also study the configuration of singular fibers under the stable degeneration limit. In F-theory, non-Abelian gauge symmetries on the 7-branes are in correspondence with the types of singular fibers. Therefore, analyzing the configurations of singular fibers under the stable degeneration limit is of physical interest. 
\par The outline of this study is as follows. In section \ref{sec2}, we discuss the stable degeneration limit where a K3 surface degenerates into two isomorphic rational elliptic surfaces. We provide an equation that systematically describes this process. We determine that any pair of isomorphic rational elliptic surfaces glued together deforms to a K3 surface, which is a double cover of $\P^2$ ramified over a sextic curve. Moreover, we discuss the configurations of singular fibers under the degeneration. Furthermore, we review the properties of the extremal rational elliptic surfaces. We discuss some examples of stable degeneration using an extremal rational elliptic surface. The sum of two isomorphic rational elliptic surfaces glued together admits a deformation to a K3 surface, the singular fibers of which are twice that of the rational elliptic surface. For special situations, two fibers of the same type of the resulting K3 surface collide, and they are enhanced to a fiber of another type. These situations can be considered as special cases of stable degeneration. For such situations, some K3 surfaces become attractive. In section \ref{subsec2.4}, we determine the complex structures of these attractive K3 surfaces and their Weierstrass forms. We also deduce the non-Abelian gauge symmetries that form on the 7-branes in F-theory compactifications on these attractive K3 surfaces times a K3. $E_6$, $E_7$, $E_8$, $SU(5)$, and $SO(10)$ gauge groups arise in these models. In section \ref{sec3}, we investigate the stable degeneration limit where a K3 surface degenerates into two non-isomorphic rational elliptic surfaces. Whether such a degeneration exists can be determined by studying the lattice structure of the second integral cohomology group $H^2(S, \Z)$ of a K3 surface $S$. We obtain the lattice condition under which two non-isomorphic rational elliptic surfaces glued together admits a deformation to a K3 surface. We determine whether a deformation to a K3 surface exists for pairs of extremal rational elliptic surfaces. We study the configuration of singular fibers under stable degeneration using the lattice theoretic argument. We state the concluding remarks in section \ref{sec4}.

\section{Stable degeneration of K3 to a pair of isomorphic rational elliptic surfaces}
\label{sec2}

\subsection{Equation for the degeneration of K3 to a pair of isomorphic rational elliptic surfaces}
\label{subsec2.1}
In this section, we discuss the stable degeneration of a K3 surface to two isomorphic rational elliptic surfaces. We provide an equation that describes such stable degeneration systematically. From this equation, we determine that any pair of isomorphic rational elliptic surfaces admits stable degeneration (i.e., there is a K3 surface that splits into that pair of rational elliptic surfaces in the stable degeneration limit). We also conclude that there always exists some appropriate K3 surface whose singular fibers are the sum of the singular fibers of the two isomorphic rational elliptic surfaces to which K3 surface degenerates. 
\par When we discuss a rational elliptic surface in this study, we only consider such a surface with a section. The base space of a rational elliptic surface is isomorphic to $\P^1$ by L\"{u}roth's theorem, and it is known that every rational elliptic surface with a section is the blow-up of $\P^2$ in the nine base points of a cubic pencil \cite{CD}. The Picard number of a rational elliptic surface is 10, and the rank of the Mordell-Weil group ranges from 0 to 8:
\begin{equation}
0 \le \rm{rk}\, MW \le 8.
\end{equation}
A generic rational elliptic surface has Mordell-Weil rank 8. Rational elliptic surfaces with Mordell-Weil rank 0 are called {\it extremal} rational elliptic surfaces. For extremal rational elliptic surfaces, the rank of the singularity type is 8, which is the highest for a rational elliptic surface. We will review the properties of extremal rational elliptic surfaces in section \ref{subsec2.2}. In section \ref{sec3}, we focus on extremal rational elliptic surfaces to discuss the stable degeneration of a K3 surface into a pair of non-isomorphic rational elliptic surfaces. 
\par When two rational elliptic surfaces have isomorphic smooth elliptic fibers, we can glue two rational elliptic surfaces in the following fashion: we choose a point in base $\P^1$ of each rational elliptic surface over which the fiber is smooth and isomorphic, and we glue two rational elliptic surfaces by identifying the isomorphic smooth fibers over the chosen points. As described in section \ref{sec3}, when a certain lattice theoretic condition is satisfied, the sum of two (not necessarily isomorphic) rational elliptic surfaces glued together can be deformed to a K3 surface. 
\par In this section, we particularly consider the case wherein two rational elliptic surfaces that are glued together are isomorphic. For this particular case, we explicitly provide an equation that describes the stable degeneration of a K3 surface into the pair of isomorphic rational elliptic surfaces glued together. As stated above, a rational elliptic surface $X$ is the blow-up of $\P^2$ in the base points of a cubic pencil; we denote the cubic pencil of a rational elliptic surface $X$ as $f$. The double cover of $\P^2$ ramified along a degree 6 curve is, in general, a K3 surface. We particularly consider double covers of $\P^2$ ramified over a degree 6 curve given by the following form of equations:
\begin{equation}
\tau^2=fg,
\label{double cover in sec2}
\end{equation}
where $g$ is a polynomial of degree 3 and $g$ is the cubic pencil of the same type as the pencil $f$, but the ratio of the coefficients of the pencil $g$ is generally different from the ratio of the coefficients of the pencil $f$. For this situation, K3 surface (\ref{double cover in sec2}) is elliptically fibered. To be explicit, when the cubic pencil $f$ is given by 
\begin{equation}
f=a\, h_1+b\, h_2,
\end{equation}
where $a, b$ are coefficients of the pencil $f$, we choose the cubic pencil $g$ as follows:
\begin{equation}
g=c\, h_1+d\, h_2.
\end{equation}
$c, d$ are coefficients of the pencil $g$, and the ratio $[c:d]$ is generally different from the ratio $[a:b]$.
For the limit at which polynomial $g$ goes to cubic pencil $f$ (i.e. for the limit at which ratio $[c:d]$ goes to the ratio $[a:b]$), equation (\ref{double cover in sec2}) is split into the following two equations:
\begin{eqnarray}
\label{two RES in sec2}
\tau & = & f \\ \nonumber
\tau & = & -f. 
\end{eqnarray}
Each of the above two equations in (\ref{two RES in sec2}) describes a rational elliptic surface given by cubic pencil $f$; therefore, when a cubic polynomial $g$ goes to cubic polynomial $f$, the K3 surface (\ref{double cover in sec2}) splits into two isomorphic rational elliptic surfaces, each given by the cubic pencil $f$. This is the stable degeneration limit of the K3 surface (\ref{double cover in sec2}) splitting into two copies of rational elliptic surfaces $X$. Thus, we conclude that two isomorphic rational elliptic surfaces glued along an isomorphic smooth fiber always admit a deformation to a K3 surface. 
\par We only consider rational elliptic surfaces with a global section in this note; therefore, they admit transformation into the Weierstrass form. The coefficients of the Weierstrass form depend on the coordinate of the base $\P^1$. We denote the homogeneous coordinate of the base $\P^1$ as $[u:v]$. In terms of the coordinate $[u:v]$, the stable degeneration of a K3 surface (\ref{double cover in sec2}) splitting into two isomorphic rational elliptic surfaces is described by the following equation:
\begin{equation}
\label{quadratic parameter in sec2}
\tau^2=u^2+2kuv+v^2.
\end{equation}
$k$ in equation (\ref{quadratic parameter in sec2}) denotes a parameter of deformation. $k$ varies along the deformation, and when $k$ assumes the values
\begin{equation}
k=\pm 1,
\end{equation}
equation (\ref{quadratic parameter in sec2}) splits into linear factors. This occurs when a K3 surface splits into two isomorphic rational elliptic surfaces. 
\par We observe from equation (\ref{quadratic parameter in sec2}) that the K3 surface (\ref{double cover in sec2}) is the quadratic base change of the rational elliptic surface (\ref{two RES in sec2}) into which the K3 surface splits in the stable degeneration limit, when $g$ is the cubic pencil of the same type as the pencil $f$. In other words, the Weierstrass equation of the K3 surface is obtained when some appropriate quadratic equations are substituted into variables $u,v$ in the coefficients of the Weierstrass form of a rational elliptic surface. Thus, the generic K3 surface (\ref{double cover in sec2}) that results from the deformation of two isomorphic rational elliptic surfaces (\ref{two RES in sec2}) glued together has twice the number of singular fibers as a rational elliptic surface.  
\par For special situations, singular fibers of the same type of the K3 surface, that is obtained as the quadratic base change of a rational elliptic surface, collide and they are enhanced to a fiber of another type. We discuss these situations in section \ref{subsec2.4}.
\par The aforementioned argument applies to the deformation of every pair of two isomorphic rational elliptic surfaces with a global section glued along smooth fiber to a K3 surface. 

\subsection{Extremal rational elliptic surfaces}
\label{subsec2.2}
We summarize the properties of extremal rational elliptic surfaces. In section \ref{subsec2.3}, we discuss the quadratic base change of extremal rational elliptic surfaces that ramifies only over smooth fibers. In section \ref{subsec2.4}, we discuss the limits at which singular fibers of the same type collide in the quadratic base change of extremal rational elliptic surfaces. We discuss the structures of attractive K3 surfaces that result from the quadratic base change of extremal rational elliptic surfaces in section \ref{subsec2.4}. 
\par Extremal rational elliptic surfaces have the singularity type of rank 8, and the Mordell-Weil groups only have torsion parts. Singular fiber types of the extremal rational elliptic surfaces are classified, and the complex structure of an extremal rational elliptic surface is uniquely determined by the fiber type, except for the surfaces with the fiber type $[I^*_0, \, I^*_0]$ \cite{MP}. J-invariant of fibers of an extremal rational elliptic surface with the fiber type $[I^*_0, \, I^*_0]$ is constant over the base $\P^1$, and the complex structure of a rational elliptic surface with the fiber type $[I^*_0, \, I^*_0]$ depends on the value of j. The complex structure of an extremal rational elliptic surface with the fiber type $[I^*_0, \, I^*_0]$ is fixed when the j-invariant of the fiber is chosen.  
\par Provided these facts, in this note, we denote an extremal rational elliptic surface using its fiber type as the subscript. For example, an extremal rational elliptic surface with the fiber type $[IV, \, IV^*]$ is denoted as $X_{[IV, \hspace{1mm} IV^*]}$. We simply use $n$ to represent the $I_n$ fiber, and $m^*$ to represent the $I^*_m$ fiber. Therefore, an extremal rational elliptic surface with singular fibers of type $IV^*$, $I_3$, and $I_1$ is denoted as $X_{[IV^*,  \, 3, 1]}$. We denote a surface with the fiber type $[I^*_0, I^*_0]$ as $X_{[0^*, \hspace{1mm} 0^*]}(j)$, because the complex structure of such a surface depends on the j-invariant of fibers. We list the configurations of singular fiber types of the extremal rational elliptic surfaces in Table \ref{tabextRESfibertype}. The Weierstrass forms of the extremal rational elliptic surfaces were also derived in \cite{MP}. We include the Weierstrass forms of the extremal rational elliptic surfaces in Table \ref{tabextRESfibertype}. 

\begingroup
\renewcommand{\arraystretch}{1.5}
\begin{table}[htb]
\centering
\resizebox{\columnwidth}{!}{%
  \begin{tabular}{|c|c|c|c|} \hline
$
\begin{array}{c}
\mbox{Extremal rational}\\
\mbox{elliptic surface} 
\end{array}
$ & Fiber type & $a_4$ & $a_6$ \\ \hline
$X_{[II, \hspace{1mm} II^*]}$ & $II^*$, $II$ & 0 & $uv^5$  \\ \hline
$X_{[III, \hspace{1mm} III^*]}$ & $III^*$, $III$ & $uv^3$ & 0  \\ \hline
$X_{[IV, \hspace{1mm} IV^*]}$ & $IV^*$, $IV$ & 0 & $u^2v^4$ \\ \hline
$X_{[0^*, \hspace{1mm} 0^*]}(j)$ & $I_0^*$, $I_0^*$ & $su^2v^2$ & $tu^3v^3$ \\ \hline
$X_{[II^*, \hspace{1mm} 1, 1]}$ & $II^*$ $I_1$ $I_1$ & $-3u^4$ & $2u^5v$ \\ \hline
$X_{[III^*, \hspace{1mm} 2, 1]}$ & $III^*$ $I_2$ $I_1$ & $-uv^3$ & $v^5(u-v)$ \\ \hline
$X_{[IV^*, \hspace{1mm} 3, 1]}$ & $IV^*$ $I_3$ $I_1$ & $v^3(24u-27v)$ & $v^4(16u^2-72uv+54v^2)$ \\ \hline
$X_{[4^*, \hspace{1mm} 1, 1]}$ & $I_4^*$ $I_1$ $I_1$ & $-3v^2(u^2-3v^2)$ & $uv^3(2u^2-9v^2)$ \\ \hline
$X_{[2^*, \hspace{1mm} 2, 2]}$ & $I_2^*$ $I_2$ $I_2$ & $-3uv(u-v)^2$ & $(u-v)^3(u^3+v^3)$ \\ \hline
$X_{[1^*, \hspace{1mm} 4, 1]}$ & $I^*_1$ $I_4$ $I_1$ & $-3(u-2v)^2(u^2-3v^2)$ & $u(u-2v)^3(2u^2-9v^2)$ \\ \hline
$X_{[9, 1, 1, 1]}$ & $I_9$ $I_1$ $I_1$ $I_1$ & $-3u(u^3+24v^3)$ & $2(u^6+36u^3v^3+216v^6)$ \\ \hline
$X_{[8, 2, 1, 1]}$ & $I_8$ $I_2$ $I_1$ $I_1$ & $-3(u^4+4u^2v^2+v^4)$ & $2u^6+12u^4v^2+15u^2v^4-2v^6$ \\ \hline 
$X_{[6, 3, 2, 1]}$ & $I_6$ $I_3$ $I_2$ $I_1$ & $-3(u^4+4u^3v-2uv^3+v^4)$ & $
\begin{array}{c}
2u^6+12u^5v+12u^4v^2-14u^3v^3\\
+3u^2v^4-6uv^5+2v^6
\end{array}
$ \\ \hline
$X_{[5, 5, 1, 1]}$ & $I_5$ $I_5$ $I_1$ $I_1$ & $
\begin{array}{c}
-3(u^4-12u^3v+14u^2v^2\\
+12uv^3+v^4)
\end{array}
$ & $
\begin{array}{c}
2(u^6-18u^5v+75u^4v^2\\
+75u^2v^4+18uv^5+v^6)
\end{array}
$  \\ \hline
$X_{[4, 4, 2, 2]}$ & $I_4$ $I_4$ $I_2$ $I_2$ & $-3(u^4-u^2v^2+v^4)$ & $2u^6-3u^4v^2-3u^2v^4+2v^6$ \\ \hline
$X_{[3, 3, 3, 3]}$ & $I_3$ $I_3$ $I_3$ $I_3$ & $-3(u^4-8uv^3)$ & $2(u^6+20u^3v^3-8v^6)$ \\ \hline
\end{tabular}
}
\caption{\label{tabextRESfibertype}Fiber types of the extremal rational elliptic surfaces, and coefficients $a_4,a_6$ of the Weierstrass form $y^2=x^3+a_4x+a_6$. $[u:v]$ is the homogeneous coordinate on the base $\P^1$. For surface $X_{[0^*, \hspace{1mm} 0^*]}(j)$, $s,t$ in coefficients $a_4, a_6$ of the Weierstrass form are complex numbers, $s,t\in\C$, with $4s^3+27t^2\ne 0$. $j$-invariant of fibers of surface $X_{[0^*, \hspace{1mm} 0^*]}(j)$ is a function of $s,t$.}
\end{table}  
\endgroup 

\par The cubic pencils for all the extremal rational elliptic surfaces, except the surfaces $X_{[II, \hspace{1mm} II^*]}$, $X_{[III, \hspace{1mm} III^*]}$, $X_{[IV, \hspace{1mm} IV^*]}$, $X_{[0^*, \hspace{1mm} 0^*]}(j)$, were obtained in \cite{Nar}. The cubic pencil for the extremal rational elliptic surface $X_{[IV, \hspace{1mm} IV^*]}$ is given by
\begin{equation}
a\, yz(y+z)+b\, x^3.
\end{equation}
$[x:y:z]$ represents the homogeneous coordinates on $\P^2$, and $[a:b]$ represents the homogeneous coordinate on $\P^1$. The cubic pencils of extremal rational elliptic surfaces are listed in Table \ref{tabpencil}.

\begingroup
\renewcommand{\arraystretch}{1.5}
\begin{table}[htb]
\begin{center}
  \begin{tabular}{|c|c|} \hline
Extremal rational elliptic surface & Pencil \\ \hline
$X_{[II^*, \hspace{1mm} 1, 1]}$ & $a\, \{3(x^2-y^2)z+2x^3\}+b\, z^3$  \\ \hline
$X_{[III^*, \hspace{1mm} 2, 1]}$ &  $a\, z\{y^2+x(z-2x)\}+b\, x^3$\\ \hline
$X_{[IV^*, \hspace{1mm} 3, 1]}$ &  $a\, xyz+b\, (x+y+z)^3$\\ \hline
$X_{[IV, \hspace{1mm} IV^*]}$ & $a\, yz(y+z)+b\, x^3$ \\ \hline
$X_{[4^*, \hspace{1mm} 1, 1]}$ & $a\, \{(2x+z)(xz+y^2)+x^3\}+b\, x(xz+y^2)$ \\ \hline
$X_{[2^*, \hspace{1mm} 2, 2]}$ &  $a\, (x+y)(xy+z^2)+b\, x^2y$ \\ \hline
$X_{[1^*, \hspace{1mm} 4, 1]}$ & $a\, x\{(x+y)z+y^2\}+b\, (x+y)z^2$ \\ \hline
$X_{[9, 1, 1, 1]}$ &  $a\, (x^2y+y^2z+z^2x-3xyz)+b\, xyz$ \\ \hline
$X_{[8, 2, 1, 1]}$ & $a\, \{x(xz-y^2)+y^2z\}+b\, (y+2z)(xz-y^2+z^2)$ \\ \hline 
$X_{[6, 3, 2, 1]}$ & $a\, z(x^2+xy+xz+y^2)+b\, (-x^2z+xyz+y^3)$ \\ \hline
$X_{[5, 5, 1, 1]}$ & $a\, yz(x+y+z)+b\, x(x+y)(x+z)$ \\ \hline
$X_{[4, 4, 2, 2]}$ & $a\, (x-y)(xy-z^2)+b\, xy(x+y-2z)$ \\ \hline
$X_{[3, 3, 3, 3]}$ & $a\, (x^3+y^3+z^3)+b\, xyz$ \\ \hline
\end{tabular}
\caption{\label{tabpencil}Pencils of extremal rational elliptic surfaces.}
\end{center}
\end{table}  
\endgroup 

\subsection{Example of generic deformation using an extremal rational elliptic surface}
\label{subsec2.3}
We discuss an example of results in section \ref{subsec2.1} using an extremal rational elliptic surface. We consider the surface $X_{[IV, \hspace{1mm} IV^*]}$. As stated in section \ref{subsec2.2}, the cubic pencil of surface $X_{[IV, \hspace{1mm} IV^*]}$ is given by
\begin{equation}
\label{cubic pencil in sec 2.3}
f=a\, yz(y+z)+b\, x^3.
\end{equation}
We fix $a,b$ to be non-zero constants in the cubic pencil $f$. We consider the cubic polynomial $g$ to be the pencil of the same type as the cubic pencil $f$:
\begin{equation}
g=c\, yz(y+z)+d\, x^3.
\label{perturbation pencil in sec 2.3}
\end{equation}
We fix $c,d$ to be non-zero constants. We choose $c,d$ so that the ratio $[c:d]$ is generally different from the ratio $[a:b]$ of the coefficients $a,b$ in the pencil $f$. For the cubic pencil $f$ (\ref{cubic pencil in sec 2.3}) and cubic pencil $g$ (\ref{perturbation pencil in sec 2.3}), equation 
\begin{equation}
\tau^2=fg
\label{double cover K3 in sec2.3}
\end{equation}
generically provides a K3 surface with two type $IV^*$ fibers and two type $IV$ fibers. For the limit at which ratio $[c:d]$ goes to $[a:b]$, the cubic pencil $g$ coincides with the cubic pencil $f$. In this stable degeneration limit, equation (\ref{double cover K3 in sec2.3}) is split into linear factors; accordingly, the K3 surface (\ref{double cover K3 in sec2.3}) splits into two extremal rational elliptic surfaces $ X_{[IV, \hspace{1mm} IV^*]}$.
\par We stated in section \ref{subsec2.1} that gluing together two isomorphic rational elliptic surfaces, each given by cubic pencil $f$, to obtain a K3 surface (\ref{double cover in sec2}) is equivalent to the quadratic base change of a rational elliptic surface over $\P^1$, when the cubic pencil $g$ is chosen to be the same type as the pencil $f$. We consider, as an example, the case wherein two extremal rational elliptic surfaces $X_{[IV, \hspace{1mm} IV^*]}$ are glued together to form a K3 surface, to demonstrate this explicitly. We discuss the generic situation in which the quadratic base change ramifies over smooth fibers. As given in Table \ref{tabextRESfibertype}, the Weierstrass form of $X_{[IV, \hspace{1mm} IV^*]}$ is given by
\begin{equation}
y^2=x^3+u^2v^4.
\end{equation}
Gluing together two $X_{[IV, \hspace{1mm} IV^*]}$s and deforming the resulting surface to a K3 surface is equivalent to substituting the following quadratic equations into variables $u,v$:
\begin{eqnarray}
\label{array base change in 2.3}
u=\alpha_1 \til{u}^2+\alpha_2 \til{u}\til{v}+\alpha_3\til{v}^2=\alpha_1(\til{u}-\beta_1 \til{v})(\til{u}-\beta_2 \til{v}) \\ \nonumber
v=\alpha_4 \til{u}^2+\alpha_5 \til{u}\til{v}+\alpha_6\til{v}^2=\alpha_4(\til{u}-\beta_3 \til{v})(\til{u}-\beta_4 \til{v}).
\end{eqnarray}
$\alpha_i$, $i=1, \cdots, 6$, are some constants, and the quadratic terms are split into linear factors on the right extreme hand sides in equation (\ref{array base change in 2.3}). 
(In (\ref{array base change in 2.3}), we assume that $\beta_1\ne \beta_2$ and $\beta_3\ne \beta_4$.) The resulting K3 surface has the following Weierstrass form:
\begin{equation}
\label{Weierstrass K3 IV in 2.3}
y^2=x^3+\alpha_1^2\alpha_4^4(\til{u}-\beta_1 \til{v})^2(\til{u}-\beta_2 \til{v})^2(\til{u}-\beta_3 \til{v})^4(\til{u}-\beta_4 \til{v})^4.
\end{equation}
We confirm from equation (\ref{Weierstrass K3 IV in 2.3}) that the resulting K3 surface has two type $IV^*$ fibers and two type $IV$ fibers, for generic constants $\alpha_i$, $i=1, \cdots, 6$.
\par When $\beta_1=\beta_2$, two fibers of type $IV$ collide and they are enhanced to a type $IV^*$ fiber. For this special case, we obtain a K3 surface with three $IV^*$ fibers. Situations of this kind, in which singular fibers of the same type collide in the quadratic base change of an extremal rational elliptic surface, will be discussed in section \ref{subsec2.4}.

\subsection{Attractive K3 surfaces as a special deformation of two extremal rational elliptic surfaces glued together}
\label{subsec2.4}
\subsubsection{Complex structures of attractive K3 surfaces, and gauge groups in F-theory compactifications}
\label{subsubsec2.4.1}
In section \ref{subsec2.1}, we mainly discussed the quadratic base change whose ramification occurs over smooth fibers. A smooth fiber remains smooth after the quadratic base change. When the quadratic base change is unramified over a singular fiber, we obtain two copies of that fiber. 
\par As we saw in section \ref{subsec2.1}, the quadratic base change of a rational elliptic surface to obtain a K3 surface is the reverse of a process in which a K3 surface splits into two isomorphic rational elliptic surfaces under stable degeneration. A quadratic base change generically ramifies only over smooth fibers; we obtain two copies of the singular fiber for each singular fiber after the quadratic base change. Thus, the resulting K3 surface has twice as many singular fibers as the rational elliptic surface for such a generic situation. We consider the special situation of the quadratic base change, where singular fibers of the same type collide and they are enhanced to a singular fiber of another type. We particularly consider the examples in which the resulting K3 surfaces after the quadratic base change are enhanced to attractive K3 surfaces \footnote{Following the convention of the term in \cite{M}, we refer to a K3 surface with the Picard number 20 as an attractive K3 surface in this study.}. We show in Table \ref{tabbasechange} the resulting fiber type after the collision of two singular fibers of the same type \cite{SchShio}.  
\begingroup
\begin{table}[htb]
\centering
  \begin{tabular}{|c|c|} \hline
  $
\begin{array}{c}
\mbox{Original fiber type of} \\
\mbox{a pair of identical singular fibers} \\
\end{array} $ & Resulting fiber type \\ \hline
$I_n$ & $I_{2n}$\\ 
$II$ &  $IV$\\
$III$ & $I^*_0$\\
$IV$ & $IV^*$\\ \hline
\end{tabular}
\caption{\label{tabbasechange}Resulting fibers after the collision of a pair of singular fibers of the same type.}
\end{table}
\endgroup

\par We consider stable degeneration in which two isomorphic extremal rational elliptic surfaces $X_{[II, \hspace{1mm} II^*]}$ are glued together. When two type $II$ fibers collide in the quadratic base change, we find from Table \ref{tabbasechange} that the resulting fiber has type $IV$. The resulting K3 surface has singular fibers of types $II^*$, $II^*$, and $IV$. The corresponding $ADE$ type is $E^2_8\oplus A_2$. This is an extremal K3 surface. An extremal K3 surface is an attractive elliptic K3 surface with a global section, with the Mordell-Weil group of rank 0. The extremal K3 surface with $ADE$ type $E^2_8\oplus A_2$ is discussed in \cite{Shioda2008}. We next consider the gluing of two isomorphic extremal rational elliptic surfaces $X_{[III, \hspace{1mm} III^*]}$, in which two type $III$ fibers collide. The resulting K3 surface has type $III^*$, $III^*$, and $I^*_0$ fibers. The corresponding $ADE$ type is $E^2_7\oplus D_4$. This is also an extremal K3 surface. This K3 surface is discussed in \cite{K2} as the Jacobian fibration of some K3 genus-one fibration without a section. For the gluing of two isomorphic extremal rational elliptic surfaces $X_{[IV, \hspace{1mm} IV^*]}$, when two type $IV$ fibers collide, the resulting fiber has type $IV^*$. Therefore, the resulting K3 surface has three type $IV^*$ singular fibers. The corresponding $ADE$ type is $E^3_6$. This is an extremal K3 surface, and this K3 surface is discussed in \cite{K} as the Jacobian fibration of an attractive K3 genus-one fibration without a section. 
\par For the gluing of two isomorphic extremal rational elliptic surfaces $X_{[II^*, \hspace{1mm} 1, 1]}$, we consider the situation in which two pairs of type $I_1$ fibers collide in the quadratic base change. The resulting fibers have type $I_2$; the resulting K3 surface has two singular fibers of type $II^*$, and two fibers of type $I_2$. When we consider the limit of the quadratic base change of extremal rational elliptic surface $X_{[III^*, \hspace{1mm} 2, 1]}$ at which two type $I_2$ fibers and two type $I_1$ fibers collide, the resulting K3 surface has fibers of types $III^*$, $III^*$, $I_4$, and $I_2$. The limit of quadratic base change of extremal rational elliptic surface $X_{[IV^*, \hspace{1mm} 3, 1]}$, at which two fibers of type $I_3$ and two fibers of type $I_1$ collide, gives K3 surface with singular fibers of types $IV^*$, $IV^*$, $I_6$, and $I_2$.
\par For the gluing of two extremal rational elliptic surfaces $X_{[4^*, \hspace{1mm} 1, 1]}$, when we consider the limit of the quadratic base change at which two pairs of type $I_1$ fibers collide, the resulting K3 surface has two singular fibers of type $I^*_4$ and two singular fibers of type $I_2$. For extremal rational elliptic surface $X_{[2^*, \hspace{1mm} 2, 2]}$, when we consider the limit of the quadratic base change at which two pairs of type $I_2$ fibers collide, the resulting K3 surface has two type $I^*_2$ fibers and two type $I_4$ fibers. For the gluing of two extremal rational elliptic surface $X_{[1^*, \hspace{1mm} 4, 1]}$, when we consider the limit of the quadratic base change at which two type $I_4$ fibers and two type $I_2$ fibers collide, the resulting K3 surface has type $I^*_1$, $I^*_1$, $I_8$, and $I_2$ fibers. 
\par For extremal rational elliptic surface $X_{[5, 5, 1, 1]}$, when we consider the limit of the quadratic base change at which two fibers of type $I_5$ and two fibers of type $I_1$ collide, the resulting K3 surface has 1 type $I_{10}$ fiber, 2 type $I_5$ fibers, 1 type $I_2$ fiber, and 2 type $I_1$ fibers. 
\par From $ADE$ types of the 10 K3 surfaces that we obtained above as the quadratic base change of extremal rational elliptic surfaces, we conclude that they are extremal K3 surfaces. $ADE$ types of the extremal K3 surfaces and the corresponding complex structures were classified in \cite{SZ}. Mordell-Weil groups of extremal K3 surfaces were also derived in \cite{SZ}. Using Table 2 in \cite{SZ}, we can deduce the complex structures of the 10 extremal K3 surfaces from their $ADE$ types. 
\par The complex structure of an attractive K3 surface $S$ is specified by the transcendental lattice $T(S)$, which is defined to be the orthogonal complement of the N\'eron--Severi lattice in the K3 lattice $H^2(S, \Z)$ \cite{SI}. Therefore, we represent the complex structure of an attractive K3 surface by the intersection matrix of the transcendental lattice in this study. For an attractive K3 surface, the transcendental lattice is a 2 $\times$ 2 integral, symmetric, positive-definite even lattice. See section 4 in \cite{K} for a review of the correspondence of the complex structures of attractive K3 surfaces and the transcendental lattice. 
\par We list the fiber types, corresponding $ADE$ types, the complex structures and the Mordell-Weil groups of the 10 extremal K3 surfaces, that we obtained above as deformation of two isomorphic extremal rational elliptic surfaces glued together, in Table \ref{tabattractiveK3}. The global structures of the non-Abelian gauge symmetries \footnote{See \cite{MV2, BIKMSV} for the correspondence of the types of the singular fibers and the non-Abelian gauge groups on the 7-branes.} that arise on the 7-branes in F-theory compactifications on the 10 extremal K3 surfaces times a K3 surface are also shown in Table \ref{tabattractiveK3}. Extremal K3 surface has the Mordell-Weil rank 0, therefore these F-theory compactifications do not have $U(1)$ gauge field.

\begingroup
\renewcommand{\arraystretch}{1.5}
\begin{table}[htb]
\begin{center}
  \begin{tabular}{|c|c|c|c|c|} \hline
Singular fibers & $ADE$ type & Complex Str. & Mordell-Weil group & Gauge group \\ \hline
$II^*$ $II^*$ $IV$ & $E_8^2\oplus A_2$ & $\begin{pmatrix}
2 & 1 \\
1 & 2 \\
\end{pmatrix}$ & 0 & $E_8^2 \times SU(3)$ \\ \hline
$III^*$ $III^*$ $I^*_0$ & $E^2_7 \oplus D_4$ & $\begin{pmatrix}
2 & 0 \\
0 & 2 \\
\end{pmatrix}$ & $\Z_2$ & $E^2_7\times SO(8) / \Z_2$ \\ \hline
$IV^*$ $IV^*$ $IV^*$ & $E^3_6$ & $\begin{pmatrix}
2 & 1 \\
1 & 2 \\
\end{pmatrix}$ & $\Z_3$ & $E^3_6 / \Z_3$ \\ \hline
$II^*$ $II^*$ $I_2$ $I_2$ & $E^2_8\oplus A_1^2$ & $\begin{pmatrix}
2 & 0 \\
0 & 2 \\
\end{pmatrix}$ & 0 & $E^2_8\times SU(2)^2$ \\ \hline
$III^*$ $III^*$ $I_4$ $I_2$ & $E^2_7\oplus A_3\oplus A_1$ & $\begin{pmatrix}
4 & 0 \\
0 & 2 \\
\end{pmatrix}$ & $\Z_2$ & $E^2_7 \times SU(4) \times SU(2) / \Z_2$ \\ \hline
$IV^*$ $IV^*$ $I_6$ $I_2$ & $E^2_6\oplus A_5 \oplus A_1$ & $\begin{pmatrix}
6 & 0 \\
0 & 2 \\
\end{pmatrix}$ & $\Z_3$ & $E_6^2\times SU(6) \times SU(2) / \Z_3$ \\ \hline
$I^*_4$ $I^*_4$ $I_2$ $I_2$ & $D_8^2\oplus A_1^2$ & $\begin{pmatrix}
2 & 0 \\
0 & 2 \\
\end{pmatrix}$ & $\Z_2\times\Z_2$ & $SO(16)^2\times SU(2)^2 / \Z_2\times\Z_2$ \\ \hline
$I^*_2$ $I^*_2$ $I_4$ $I_4$ & $D^2_6\oplus A^2_3$ & $\begin{pmatrix}
4 & 0 \\
0 & 4 \\
\end{pmatrix}$ & $\Z_2\times\Z_2$ & $SO(12)^2\times SU(4)^2 / \Z_2\times\Z_2$ \\ \hline
$I^*_1$ $I^*_1$ $I_8$ $I_2$ & $D^2_5\oplus A_7\oplus A_1$ & $\begin{pmatrix}
8 & 0 \\
0 & 2 \\
\end{pmatrix}$ & $\Z_4$ & $SO(10)^2\times SU(8)\times SU(2) / \Z_4$ \\ \hline
$I_{10}$ $I_5$ $I_5$ $I_2$ $I_1$ $I_1$ & $A_9\oplus A_4^2\oplus A_1$ & $\begin{pmatrix}
10 & 0 \\
0 & 2 \\
\end{pmatrix}$ & $\Z_5$ & $SU(10)\times SU(5)^2\times SU(2) / \Z_5$ \\ \hline
\end{tabular}
\caption{\label{tabattractiveK3}Configurations of singular fibers, $ADE$ types, complex structures and the Mordell-Weil groups of 10 extremal K3 surfaces. We also list the global structures of the non-Abelian gauge groups that form on the 7-branes.}
\end{center}
\end{table}  
\endgroup 

\subsubsection{Weierstrass equation}
We obtained 10 extremal K3 surfaces in section \ref{subsubsec2.4.1} as the limit of the quadratic base change \footnote{The relationship of K3 surfaces and rational elliptic surfaces via base change is also discussed in \cite{Sch2007}.} of extremal rational elliptic surfaces, at which singular fibers of the same type collide. Therefore, by substituting appropriate quadratic polynomials into $u,v$ in the Weierstrass forms of extremal rational elliptic surfaces, listed in Table \ref{tabextRESfibertype} in section \ref{subsec2.2}, we can deduce the Weierstrass forms of the 10 extremal K3 surfaces, obtained in section \ref{subsubsec2.4.1}. 
\par As an example, extremal K3 surface whose transcendental lattice has the intersection matrix $\begin{pmatrix}
6 & 0 \\
0 & 2 \\
\end{pmatrix}$, with $ADE$ type $E^2_6\oplus A_5\oplus A_1$, is obtained via the quadratic base change of extremal rational elliptic surface $X_{[IV^*, \hspace{1mm} 3, 1]}$, in which two fibers of type $I_3$ and two fibers of type $I_1$ collide. The Weierstrass form of extremal rational elliptic surface $X_{[IV^*, \hspace{1mm} 3, 1]}$ is given by
\begin{equation}
\label{Weierstrass431 in 2.4.2}
y^2=x^3+v^3(24u-27v)x+v^4(16u^2-72uv+54v^2).
\end{equation}
The discriminant of the Weierstrass form (\ref{Weierstrass431 in 2.4.2}) is \cite{MP}
\begin{equation}
\Delta \sim u^3v^8(u-v).
\end{equation}
Type $IV^*$ fiber is at $[u:v]=[1:0]$, type $I_3$ fiber is at $[u:v]=[0:1]$ and type $I_1$ fiber is at $[u:v]=[1:1]$. We consider the following substitutions for $u,v$ in the Weierstrass form of extremal rational elliptic surface $X_{[IV^*, \hspace{1mm} 3, 1]}$:
\begin{eqnarray}
u & = & \til{u}^2 \\ \nonumber
v & = & 2\til{u}\til{v}-\til{v}^2.
\end{eqnarray}
This gives the limit of the quadratic base change at which two fibers of type $I_3$ collide at $[u:v]=[0:1]$ and two fibers of type $I_1$ collide at $[u:v]=[1:1]$. The resulting equation 
\begin{equation}
\begin{split}
y^2= & \, x^3+v^3(2u-v)^3 (24u^2+27v^2-54uv)x \\
         & +v^4(2u-v)^4(16u^4-144u^3v+288u^2v^2-216uv^3+54v^4)
\end{split}
\label{WeierstrassK3 in 2.4.2}
\end{equation}
gives the Weierstrass form of extremal K3 surface with transcendental lattice $\begin{pmatrix}
6 & 0 \\
0 & 2 \\
\end{pmatrix}$, with $ADE$ type $E^2_6\oplus A_5\oplus A_1$. The discriminant of the Weierstrass form (\ref{WeierstrassK3 in 2.4.2}) is given by
\begin{equation}
\Delta \sim u^6v^8(2u-v)^8(u-v)^2.
\label{discK3 in 2.4.2}
\end{equation}
We confirm from equations (\ref{WeierstrassK3 in 2.4.2}) and (\ref{discK3 in 2.4.2}) that extremal K3 surface (\ref{WeierstrassK3 in 2.4.2}) has two type $IV^*$ fibers, at $[u:v]=[1:0], [1:2]$, type $I_6$ fiber at $[u:v]=[0:1]$ and type $I_2$ fiber at $[u:v]=[1:1]$. 
\par We show the Weierstrass forms of the 10 extremal K3 surfaces, which we obtained in section \ref{subsubsec2.4.1}, in Table \ref{tabWeierstrassK3}. The discriminants of the Weierstrass forms in Table \ref{tabWeierstrassK3} are listed in Table \ref{tabdiscK3}. The Weierstrass forms of extremal K3 surfaces with $ADE$ types $E^2_8\oplus A_2$ and $E^2_8\oplus A_1^2$ are discussed in \cite{Shioda2008}. The Weierstrass forms of extremal K3 surfaces with $ADE$ types $E^3_6$ and $E^2_7\oplus D_4$ are discussed in \cite{K} and \cite{K2}, respectively. 

\begingroup
\renewcommand{\arraystretch}{1.5}
\begin{table}[htb]
\centering
\resizebox{\columnwidth}{!}{%
  \begin{tabular}{|c|c|c|} \hline
$ADE$ type & $f$ & $g$  \\ \hline
$E_8^2\oplus A_2$ & 0 & $(u-v)^2 u^5 v^5$     \\ \hline
$E^2_7 \oplus D_4$ & $(u-v)^2 u^3 v^3$ & 0    \\ \hline
$E^3_6$ & 0 & $(u-v)^4 u^4 v^4$   \\ \hline
$E^2_8\oplus A_1^2$ & $-3\, u^4 v^4$ & $u^5 v^5 (u^2+v^2)$   \\ \hline
$E^2_7\oplus A_3\oplus A_1$ & $-\frac{9}{16} \, (u^2+v^2+\frac{10}{3}uv) u^3v^3$ & $\frac{9}{4}\, u^5v^5 (\frac{1}{4}u^2+\frac{1}{4}v^2+\frac{7}{18}uv)$  \\ \hline
$E^2_6\oplus A_5 \oplus A_1$ & $v^3(2u-v)^3 (24u^2+27v^2-54uv)$ & $
\begin{array}{c}
v^4(2u-v)^4
\cdot (16u^4-144u^3v
\\+288u^2v^2-216uv^3+54v^4)
\end{array}
$  \\ \hline
$D_8^2\oplus A_1^2$ & $-3u^2v^2(u^4+v^4-u^2v^2)$ & $(u^2+v^2)u^3v^3(2u^4-5u^2v^2+2v^4)$   \\ \hline
$D^2_6\oplus A^2_3$ & $
\begin{array}{c}
-3\cdot \frac{4}{\omega(\omega-1)}uv[u^2+v^2+(2+\frac{4}{\omega-1})uv]\\
\cdot[u^2+v^2+(2+\frac{4}{\omega})uv]^2
\end{array}
$  & $
\begin{array}{c}
[u^2+v^2+(2+\frac{4}{\omega})uv]^3\\
\cdot\{[u^2+v^2+(2+\frac{4}{\omega-1})uv]^3+(\frac{4}{\omega-1}uv)^3\}
\end{array}
$  \\ \hline
$D^2_5\oplus A_7\oplus A_1$ & $
\begin{array}{c}
-3(u^4+4u^3v+4u^2v^2-\frac{3}{4}v^4)\\
\cdot(u^2+2uv-v^2)^2
\end{array}
$  & $
\begin{array}{c}
u(u+2v)(2u^4+8u^3v+8u^2v^2-\frac{9}{4}v^4)\\
\cdot(u^2+2uv-v^2)^3
\end{array}
$  \\ \hline
$A_9\oplus A_4^2\oplus A_1$ &
$
\begin{array}{c}
-3[u^8-12u^6\cdot \frac{2}{11+5\sqrt{5}}(2uv-v^2) \\
+14u^4(\frac{2}{11+5\sqrt{5}}(2uv-v^2))^2 \\
+12u^2(\frac{2}{11+5\sqrt{5}}(2uv-v^2))^3 \\
+(\frac{2}{11+5\sqrt{5}}(2uv-v^2))^4]
\end{array}
$
  &  
$
\begin{array}{c}
2[u^{12}-18u^{10}\cdot \frac{2}{11+5\sqrt{5}}(2uv-v^2) \\
+75u^8(\frac{2}{11+5\sqrt{5}}(2uv-v^2))^2 \\
+75u^4(\frac{2}{11+5\sqrt{5}}(2uv-v^2))^4 \\
+18u^2(\frac{2}{11+5\sqrt{5}}(2uv-v^2))^5 \\
+(\frac{2}{11+5\sqrt{5}}(2uv-v^2))^6]
\end{array}
$  
   \\ \hline
\end{tabular}
}
\caption{\label{tabWeierstrassK3}$ADE$ types of extremal K3 surfaces, and coefficients $f,g$ of the Weierstrass form $y^2=x^3+fx+g$. $[u:v]$ denotes the homogeneous coordinate on the base $\P^1$. $\omega$ denotes a cube root of unity, $\omega\ne 1$.}
\end{table}  
\endgroup 

\begingroup
\renewcommand{\arraystretch}{1.5}
\begin{table}[htb]
\begin{center}
  \begin{tabular}{|c|c|} \hline
$ADE$ type & $\Delta$ \\ \hline
$E_8^2\oplus A_2$ & $(u-v)^4 u^{10} v^{10}$   \\ \hline
$E^2_7 \oplus D_4$ & $(u-v)^6 u^9 v^9$   \\ \hline
$E^3_6$ & $(u-v)^8 u^8 v^8$  \\ \hline
$E^2_8\oplus A_1^2$ & $u^{10} v^{10} (u-v)^2 (u+v)^2$  \\ \hline
$E^2_7\oplus A_3\oplus A_1$ & $u^9 v^9 (u-v)^4 (u+v)^2$  \\ \hline
$E^2_6\oplus A_5 \oplus A_1$ & $u^6v^8(2u-v)^8(u-v)^2$ \\ \hline
$D_8^2\oplus A_1^2$ & $u^{10}v^{10}(u-v)^2(u+v)^2$  \\ \hline
$D^2_6\oplus A^2_3$ & $(u-v)^4(u+v)^4[u^2+v^2+(2+\frac{4}{\omega})uv]^8$  \\ \hline
$D^2_5\oplus A_7\oplus A_1$ & $v^8(u^2+2uv-v^2)^7(u+v)^2$ \\ \hline
$A_9\oplus A_4^2\oplus A_1$ & $
\begin{array}{c}
u^{10}v^5(v-2u)^5(u-v)^2 \\
\cdot [u^2-\frac{11-5\sqrt{5}}{11+5\sqrt{5}}(2uv-v^2)]
\end{array}
$  \\ \hline
\end{tabular}
\caption{\label{tabdiscK3}Discriminant $\Delta$ of the Weierstrass forms in Table \ref{tabWeierstrassK3} are shown. We suppressed the irrelevant constant factors of the discriminants.}
\end{center}
\end{table}  
\endgroup 

\subsubsection{Anomaly cancellation condition}
We consider F-theory compactifications on the extremal K3 surfaces obtained in section \ref{subsubsec2.4.1} times a K3 surface. The resulting theory is a four-dimensional theory with $N=2$, without a four-form flux. The anomaly cancellation condition determines the form of the discriminant locus to be 24 K3 surfaces, counted with multiplicity; there are 24 7-branes wrapped on the K3 surfaces \footnote{See, for example, \cite{K} for discussion.}. We show the correspondence of the numbers of 7-branes and the fiber types in Table \ref{tabnumber7-branes}. 
\begingroup
\renewcommand{\arraystretch}{1.1}
\begin{table}[htb]
\centering
  \begin{tabular}{|c|c|} \hline
Fiber type & \# of 7-branes (Euler number) \\ \hline
$I_n$ & $n$\\
$I^*_0$ &  6\\ 
$I^*_m$ &  $m+$6\\ 
$II$ &  2\\
$III$ & 3\\
$IV$ & 4\\
$IV^*$ & 8\\ 
$III^*$ & 9\\
$II^*$ & 10\\ \hline
\end{tabular}
\caption{\label{tabnumber7-branes}Associated numbers of 7-branes for fiber types.}
\end{table}
\endgroup
The Euler numbers of the singular fibers can be found in \cite{Kod2}. The Euler number of fiber type can be considered as the number of the associated 7-branes. We confirm from Tables \ref{tabattractiveK3} and \ref{tabnumber7-branes} that there are in fact 24 7-branes in F-theory compactifications on the 10 extremal K3 surfaces times a K3 surface. Therefore, we conclude that the anomaly cancellation condition is satisfied for these compactifications. 
\par By turning on four-form flux \cite{BB, SVW, W, GVW, DRS}, F-theory compactification on K3 times K3 gives four-dimensional theory with $N=1$ supersymmetry. We confirm from Table 1 in \cite{AK} and Table 2 in \cite{BKW} that for F-theory compactifications on the 10 extremal K3 surfaces, obtained in section \ref{subsubsec2.4.1}, times some appropriate attractive K3 surface, the tadpole \cite{SVW} can be cancelled. See \cite{K, K2} for the details.

\section{Stable degeneration of K3 to a pair of non-isomorphic rational elliptic surfaces}
\label{sec3}
In this section, we discuss the stable degeneration where a K3 surface degenerates into two non-isomorphic rational elliptic surfaces. It is considerably difficult to provide a general equation to describe this kind of stable degeneration. Therefore, instead of providing an equation to describe the process, we use a lattice theoretic approach to determine whether stable degeneration exists for pairs of non-isomorphic rational elliptic surfaces. For the sake of brevity, in this section, we simply say ``a pair of rational elliptic surfaces'' to indicate a pair of non-isomorphic rational elliptic surfaces. We also discuss the configurations of singular fibers under stable degeneration. In section \ref{subsec3.1}, we will discuss a lattice theoretic condition for the existence of stable degeneration for pairs of rational elliptic surfaces. Applying the lattice condition, in section \ref{subsec3.2}, we demonstrate that stable degeneration exists for pairs of extremal rational elliptic surfaces. 
\par Elliptic fiber of a rational elliptic surface generally has the moduli of dimension 1 over the base $\P^1$. Therefore, given a pair of rational elliptic surfaces, there is a pair of isomorphic smooth elliptic fibers, and the pair of rational elliptic surfaces can be glued along the isomorphic smooth fibers. However, when elliptic fiber of a rational elliptic surface has the constant moduli over the base $\P^1$, such gluing is not necessarily possible. For the three extremal rational elliptic surfaces $X_{[II, \hspace{1mm} II^*]}$, $X_{[III, \hspace{1mm} III^*]}$, $X_{[IV, \hspace{1mm} IV^*]}$, the complex structure of elliptic fibers is constant over the base $\P^1$. \footnote{The complex structure of elliptic fibers of extremal rational elliptic surface $X_{[0^*, \hspace{1mm} 0^*]}(j)$ is also constant over the base $\P^1$, but there is a degree of freedom in choosing j-invariant of an elliptic fiber. Therefore, when j-invariant is appropriately chosen, $X_{[0^*, \hspace{1mm} 0^*]}(j)$ can be glued with another rational elliptic surface. For this reason, we include extremal rational elliptic surface $X_{[0^*, \hspace{1mm} 0^*]}(j)$.} We do not consider the three extremal rational elliptic surfaces $X_{[II, \hspace{1mm} II^*]}$, $X_{[III, \hspace{1mm} III^*]}$, $X_{[IV, \hspace{1mm} IV^*]}$ in this section.

\subsection{Lattice condition for stable degeneration limit}
\label{subsec3.1}
We use the Torelli theorem for K3 surfaces \cite{PS-S} to deduce a lattice theoretic condition that determines whether pairs of rational elliptic surfaces admit stable degeneration. The Torelli theorem for K3 surfaces states that the geometry of a K3 surface is determined by the structure of the K3 lattice $\Lambda_{\rm K3}$. The K3 lattice $\Lambda_{\rm K3}$ of a K3 surface $S$ is the second integral cohomology group $H^2(S, \Z)$. The K3 lattice $\Lambda_{\rm K3}$ is the indefinite even unimodular lattice of signature (3,19), and it is the direct sum of three copies of the hyperbolic plane $U$ and two copies of the $E_8$ lattice:
\begin{equation}
\Lambda_{\rm K3} \cong U^3\oplus E^2_8.
\end{equation}
In this note, we assume that a rational elliptic surface has a global section; thus, we presume that the K3 surface obtained as a deformation of the sum of two rational elliptic surfaces also admits a global section \footnote{In general, genus-one fibered K3 surfaces need not have a global section. For discussion of the geometry of genus-one fibered K3 surfaces without a section and string compactifications on such spaces, see, for example, \cite{MTsection, K, K2}. For recent progress in F-theory compactifications on genus-one fibrations without a section, see, for example, also \cite{BM, AGGK, KMOPR, GGK, MPTW, MPTW2, BGKintfiber, CDKPP, LMTW, KCY4, Kdisc}.}. An elliptic K3 surface having a section is equivalent to the primitive embedding of the hyperbolic plane $U$ into the K3 lattice $\Lambda_{\rm K3}$ \cite{PS-S, Kondo1}. We denote the $ADE$ types of two rational elliptic surfaces, $X_1$ and $X_2$, as $R_1$ and $R_2$, respectively. Applying the argument in \cite{Kondo2} to the boundary of the closure of K3 moduli, from the Torelli theorem for K3 surfaces, we deduce that a K3 surface exists and it admits stable degeneration into two rational elliptic surfaces $X_1$ and $X_2$, exactly when there is a primitive embedding of the lattice $U\oplus R_1\oplus R_2$ into the K3 lattice $\Lambda_{\rm K3}$:
\begin{equation}
U\oplus R_1\oplus R_2 \subset \Lambda_{\rm K3}.
\end{equation}
\par Whether lattice $U\oplus R_1\oplus R_2$ primitively embeds into the K3 lattice $\Lambda_{\rm K3}$ can be determined by the criterion given in \cite{Nik}. Some lattice theoretic terms are necessary to state the criterion; we introduce some lattice theoretic terms first. 
\par By lattice, we indicate a finite rank free $\Z$-module with a non-degenerate integral symmetric bilinear form. The lattice $L$ is said to be even, when for every element $x$ of $L$, $x^2=x\cdot x$ is even. The discriminant of lattice $L$, disc $L$, is the determinant of an intersection matrix $(e_i\cdot e_j)_{ij}$ for a basis $\{e_i\}$ of the lattice $L$. The lattice $L$ is said to be unimodular when its discriminant is $\pm 1$. $U$ denotes the hyperbolic plane. The hyperbolic plane $U$ is the even unimodular lattice of signature (1,1). $E_8$ denotes the even unimodular lattice of signature (0,8). $E_8$ and $U$ are unique up to the isometries of lattice. When the lattice $L_1$ embeds into the lattice $L_2$, the embedding $L_1\subset L_2$ is said to be primitive when the quotient $L_2/L_1$ is free as a $\Z$-module. The dual lattice of lattice $L$ is the lattice $Hom(L,\Z)$, and is denoted as $L^*$. The quotient $G_L:=L^*/L$ is a finite Abelian group, and this group is called the {\it discriminant group}. When $L$ is an even lattice, the map $q_L: G_L \rightarrow \Q/2\Z$, 
\begin{equation}
q_L(x)=x^2 \hspace{2mm} mod \hspace{1mm} 2\Z
\end{equation}
defines a non-degenerate quadratic form of the discriminant group $G_L$; form $q_L$ is called the {\it discriminant form}.
\par When $A$ is a finite Abelian group, its length $l(A)$ is defined to be the minimum number of elements required to generate group $A$. Let $(A, q)$ be a pair of the finite Abelian group $A$ and the quadratic form $q$ defined on the group $A$. 
\par Further, we state the criterion of primitive lattice embedding. 

\vspace{5mm}

\noindent Criterion ($C$) \cite{Nik} \\
$M$ is an even lattice of signature ($m_+, m_-$), $G$ is the discriminant group of $M$ and $q$ is the discriminant form of $M$. Then, $M$ primitively embeds into some even unimodular lattice of signature ($l_+, l_-$) when all of the three conditions i) - iii) are satisfied: \\
i) $l_++l_--{\rm rk} \, M > l(G)$. \\
ii) $l_+-m_+\ge 0$, \hspace{2mm} $l_--m_-\ge 0$. \\
iii) $l_+-l_-\equiv 0$ (mod 8).

\subsection{Pairs of extremal rational elliptic surfaces}
\label{subsec3.2}
Applying criterion ($C$), we discuss whether the lattice $U\oplus R_1\oplus R_2$ of pairs of extremal rational elliptic surfaces, $X_1\amalg X_2$, primitively embeds into the K3 lattice $\Lambda_{\rm K3}$. This determines whether stable degeneration exists for pairs of extremal rational elliptic surfaces. An even unimodular lattice of signature (3,19) is unique up to the isometries of lattice \cite{Mil}. Therefore, when the lattice $U\oplus R_1\oplus R_2$ admits a primitive embedding into some even unimodular lattice of signature (3,19), it primitively embeds into the K3 lattice $\Lambda_{\rm K3}$.
\par As stated in section \ref{subsec2.2}, the complex structures and singular fiber types of the extremal rational elliptic surfaces were classified. For an extremal rational elliptic surface, the discriminant group and the Mordell-Weil group are identical \cite{Shioda}. The Mordell-Weil groups of the rational elliptic surfaces were computed in \cite{OS}. We list the discriminant groups and $ADE$ types of extremal rational elliptic surfaces in Table \ref{tabdiscriminant}.

\begingroup
\renewcommand{\arraystretch}{1.5}
\begin{table}[htb]
\begin{center}
  \begin{tabular}{|c|c|c|} \hline
Extremal rational elliptic surface & $ADE$ type & Discriminant group \\ \hline
$X_{[0^*, \hspace{1mm} 0^*]}(j)$ & $D_4^2$ & $\Z/2\Z\times\Z/2\Z$ \\ \hline
$X_{[II^*, \hspace{1mm} 1, 1]}$ & $E_8$ & $0$ \\ \hline
$X_{[III^*, \hspace{1mm} 2, 1]}$ & $E_7\oplus A_1$ & $\Z/2\Z$ \\ \hline
$X_{[IV^*, \hspace{1mm} 3, 1]}$ & $E_6\oplus A_2$ & $\Z/3\Z$ \\ \hline
$X_{[4^*, \hspace{1mm} 1, 1]}$ & $D_8$ & $\Z/2\Z$ \\ \hline
$X_{[2^*, \hspace{1mm} 2, 2]}$ & $D_6\oplus A_1^2$ & $\Z/2\Z\times\Z/2\Z$ \\ \hline
$X_{[1^*, \hspace{1mm} 4, 1]}$ & $D_5\oplus A_3$ & $\Z/4\Z$ \\ \hline
$X_{[9, 1, 1, 1]}$ & $A_8$ & $\Z/3\Z$ \\ \hline
$X_{[8, 2, 1, 1]}$ & $A_7\oplus A_1$ & $\Z/4\Z$ \\ \hline 
$X_{[6, 3, 2, 1]}$ & $A_5\oplus A_2\oplus A_1$ & $\Z/6\Z$ \\ \hline
$X_{[5, 5, 1, 1]}$ & $A_4^2$ & $\Z/5\Z$ \\ \hline
$X_{[4, 4, 2, 2]}$ & $A_3^2\oplus A_1^2$ & $\Z/4\Z\times\Z/2\Z$ \\ \hline
$X_{[3, 3, 3, 3]}$ & $A_2^4$ & $\Z/3\Z\times\Z/3\Z$ \\ \hline
\end{tabular}
\caption{\label{tabdiscriminant} The discriminant groups and $ADE$ types of the extremal rational elliptic surfaces.}
\end{center}
\end{table}  
\endgroup 

\par Let $G$ denote the product group of the discriminant groups of two extremal rational elliptic surfaces $X_1$ and $X_2$. We determine the pairs of extremal rational elliptic surfaces that satisfy the condition i) 
\begin{equation}
{\rm rk} \, \Lambda_{\rm K3}-{\rm rk} \, U\oplus R_1\oplus R_2 > l(G).
\end{equation}
in criterion ($C$). 
From Table \ref{tabdiscriminant}, we observe that the length of the discriminant group of an extremal rational elliptic surface is either 1 or 2. For two extremal rational elliptic surfaces, $X_1$ and $X_2$, the singular fiber types $R_1$ and $R_2$ are rank 8 even lattices
\begin{equation}
{\rm rk} \, R_1={\rm rk} \, R_2=8
\end{equation}
of signature (0,8). Thus, the lattice $U\oplus R_1\oplus R_2$ has the signature (1,17). The difference of the rank of the K3 lattice $\Lambda_{\rm K3}$ and the rank of the lattice $U\oplus R_1\oplus R_2$ is
\begin{equation}
3+19-(1+17)=4.
\end{equation}
The length $l(G)$ of the discriminant group $G$ attains this bound only when the discriminant groups of two extremal rational elliptic surfaces both have the length 2. The product group $G$ of the discriminant groups of extremal rational elliptic surfaces has the length 4 only for three pairs \footnote{The product of discriminant group $\Z/3\Z\times \Z/3\Z$ and another discriminant group with length 2 has length 2. For example, the product of $\Z/3\Z\times\Z/3\Z$ with $\Z/2\Z\times \Z/2\Z$ is isomorphic to $\Z/6\Z\times \Z/6\Z$, which has the length 2.} of extremal rational elliptic surfaces:
\begin{equation}
X_{[4, 4, 2, 2]}\amalg X_{[2^*, \hspace{1mm} 2, 2]}, \hspace{2mm} X_{[2^*, \hspace{1mm} 2, 2]}\amalg X_{[0^*, \hspace{1mm} 0^*]}, \hspace{2mm} X_{[0^*, \hspace{1mm} 0^*]}\amalg X_{[4, 4, 2, 2]}.
\end{equation}
For all other pairs of extremal rational elliptic surfaces, the criterion ($C$) applies. Both conditions ii) and iii) in criterion ($C$) are satisfied:
\begin{equation}
3-1>0, \hspace{1mm} 19-17>0,
\end{equation}
and
\begin{equation}
3-19 \equiv 0 \hspace{2mm} (mod \hspace{1mm} 8).
\end{equation}
Thus, we determine that the lattice $U\oplus R_1\oplus R_2$ primitively embeds into the K3 lattice $\Lambda_{\rm K3}$ for all pairs of extremal rational elliptic surfaces, except the three pairs $X_{[4, 4, 2, 2]}\amalg X_{[2^*, \hspace{1mm} 2, 2]}$, $X_{[2^*, \hspace{1mm} 2, 2]}\amalg X_{[0^*, \hspace{1mm} 0^*]}$ and $X_{[0^*, \hspace{1mm} 0^*]}\amalg X_{[4, 4, 2, 2]}$. We conclude that the stable degeneration of a K3 surface exists for all pairs of extremal rational elliptic surfaces except the three pairs. 
\par The remaining three pairs, $X_{[4, 4, 2, 2]}\amalg X_{[2^*, \hspace{1mm} 2, 2]}$, $X_{[2^*, \hspace{1mm} 2, 2]}\amalg X_{[0^*, \hspace{1mm} 0^*]}$ and $X_{[0^*, \hspace{1mm} 0^*]}\amalg X_{[4, 4, 2, 2]}$, have the $ADE$ types $D_6\oplus A^2_3\oplus A^4_1$, $D_6\oplus D^2_4\oplus A^2_1$, $D^2_4\oplus A^2_3\oplus A^2_1$, respectively. $ADE$ types of the singular fibers of elliptic K3 surfaces with a global section were classified in \cite{Shimada}. We conclude from Table 1 in \cite{Shimada} that the lattice $U\oplus R_1\oplus R_2$ primitively embeds into the K3 lattice $\Lambda_{\rm K3}$ for the three pairs of extremal rational elliptic surfaces. (They correspond to No.2079, 2043, 2152 in Table 1 in \cite{Shimada}, respectively.) This demonstrates that the stable degeneration of a K3 surface exists for the remaining three pairs of extremal rational elliptic surfaces.
\par The aforementioned argument demonstrates that stable degeneration exists for all pairs \footnote{As we stated at the beginning of this section, the three extremal rational elliptic surfaces $X_{[II, \hspace{1mm} II^*]}$, $X_{[III, \hspace{1mm} III^*]}$, $X_{[IV, \hspace{1mm} IV^*]}$ are not considered in this section.} of extremal rational elliptic surfaces.
\par  From lattice embedding
\begin{equation}
U\oplus R_1\oplus R_2 \subset \Lambda_{\rm K3},
\end{equation}
we deduce that the $ADE$ type of the singular fibers of the resulting K3 surface is the sum of the $ADE$ types of the singular fibers of the two non-isomorphic extremal rational elliptic surfaces.  

\section{Conclusion}
\label{sec4}
In this study, we analyzed the stable degeneration of a K3 surface into two rational elliptic surfaces. We also discussed the configurations of singular fibers under the stable degeneration limit.  
\par We demonstrated that gluing together two isomorphic rational elliptic surfaces and deforming the resulting surface to a K3 surface is always possible. We provided an equation to describe this kind of stable degeneration. The sum of two isomorphic rational elliptic surfaces glued together admits a deformation to a K3 surface, the singular fibers of which are twice the singular fibers of the rational elliptic surface. For special cases, two fibers of the same type of the resulting K3 surface collide, and they are enhanced to a fiber of another type. Some K3 surfaces become attractive in these cases. We determined the complex structures and the Weierstrass forms of these attractive K3 surfaces. We also deduced the gauge groups in F-theory compactifications on these attractive K3 surfaces times a K3.
\par We also investigated the deformation of two non-isomorphic rational elliptic surfaces glued together to a K3 surface, using the Torelli theorem of K3 surfaces. We deduced the lattice theoretic condition that must be satisfied to ensure that a deformation to a K3 surface exists for pairs of non-isomorphic rational elliptic surfaces. We confirmed that the lattice condition is satisfied for all pairs of the extremal rational elliptic surfaces. Thus, all such pairs of extremal rational elliptic surfaces glued together admit a deformation to a K3 surface. This demonstrates that for any pair of extremal rational elliptic surfaces, except the three extremal rational elliptic surfaces $X_{[II, \hspace{1mm} II^*]}$, $X_{[III, \hspace{1mm} III^*]}$, $X_{[IV, \hspace{1mm} IV^*]}$, there is a K3 surface that stably degenerates into that pair. The $ADE$ type of singular fibers of the resulting K3 surface is the sum of those of the two non-isomorphic extremal rational elliptic surfaces glued together. The lattice condition discussed in this study can be extended to general pairs of rational elliptic surfaces.

\section*{Acknowledgments}

We would like to thank Shun'ya Mizoguchi and Shigeru Mukai for discussions. This work is partially supported by Grant-in-Aid for Scientific Research {\#}16K05337 from The Ministry of Education, Culture, Sports, Science and Technology of Japan.

\end{document}